\documentclass[reprint,amsmath,amssymb,aps,prl,superscriptaddress]{revtex4-2}

\usepackage[utf8]{inputenc}
\usepackage[T1]{fontenc}
\usepackage{graphicx}
\usepackage{dcolumn}
\usepackage{bm}
\usepackage{color}
\usepackage[colorlinks=true,linkcolor=blue,urlcolor=blue,citecolor=blue,pdfusetitle]{hyperref}
\usepackage{upgreek,textgreek}
\usepackage{physics}
\usepackage{orcidlink}
\usepackage{booktabs}

\newcommand{\Hop}{\hat{H}}
\newcommand{\aop}{\hat{a}}
\newcommand{\aopd}{\hat{a}^\dag}
\newcommand{\nop}{\hat{n}}
\newcommand{\Ubb}{U_{bb}}
\newcommand{\UbI}{U_{bI}}
\newcommand{\Vho}{V_\text{ho}}
\newcommand{\lCF}{\ell_{\text{CF}}}
\newcommand{\MCF}{M_{\text{CF}}}
\newcommand{\rbI}{r_{bI}}
\newcommand{\CSF}{\mathcal{C}^{(b)}_{\text{SF}}}

\newcommand{\CAP}{\mathcal{C}_{\text{AP}}}
\newcommand{\nICF}{n_{I}^{(\text{CF})}}
\newcommand{\MI}{|\text{MI}\rangle}
\newcommand{\iIH}{|i_{\text{IH}}\rangle}
\newcommand{\PsiCF}{|\Psi_{\text{CF}}\rangle}
\newcommand{\MIc}{\langle\text{MI}|}

\newcommand{\PsiCFc}{\langle\Psi_{\text{CF}}|}

\begin{document}

\title{Counterflow of lattice polarons in harmonically confined optical lattices}

\author{Felipe Isaule~\orcidlink{0000-0003-1810-0707}}
\affiliation{Instituto de Física, Pontificia
Universidad Católica de Chile,
Avenida Vicuña Mackenna 4860,
Santiago, Chile.}

\author{Abel Rojo-Francàs~\orcidlink{0000-0002-0567-7139}}
\affiliation{Departament de F{\'i}sica Qu{\`a}ntica i Astrof{\'i}sica, Facultat de F{\'i}sica, Universitat de Barcelona, E-08028 Barcelona, Spain}
\affiliation{Institut de Ci{\`e}ncies del Cosmos, Universitat de Barcelona, ICCUB, Mart{\'i} i Franqu{\`e}s 1, E-08028 Barcelona, Spain.}

\author{Luis Morales-Molina}
\affiliation{Instituto de Física, Pontificia
Universidad Católica de Chile,
Avenida Vicuña Mackenna 4860,
Santiago, Chile.}

\author{Bruno Juliá-Díaz~\orcidlink{0000-0002-0145-6734}}
\affiliation{Departament de F{\'i}sica Qu{\`a}ntica i Astrof{\'i}sica, Facultat de F{\'i}sica, Universitat de Barcelona, E-08028 Barcelona, Spain}
\affiliation{Institut de Ci{\`e}ncies del Cosmos, Universitat de Barcelona, ICCUB, Mart{\'i} i Franqu{\`e}s 1, E-08028 Barcelona, Spain.}

\begin{abstract}
We study a mobile impurity in a one-dimensional harmonically confined optical lattice interacting repulsively with a bosonic bath. The behavior of the impurity across baths with superfluid and Mott-insulator domains is examined, including its full back-action effect on the bath. We characterize the bath-impurity phase diagram and reveal the appearance of a correlated counterflow phase,  which we support with an analytical model for a mobile impurity-hole pair. This phase shows an extended combined insulator domain of unity filling but no independent domain of constant density. The transition to this phase features a sudden orthogonality and the change of the shape of the impurity's profile to that of a free particle in an infinite square well. 
The findings of this work suggest the appearance of unconventional counterflow in trapped imbalanced atomic mixtures.

\end{abstract}

\maketitle

\textbf{\textit{Introduction.}}-- Impurities immersed in quantum mediums often form dressed quasiparticles known as polarons~\cite{landau_effective_1948,baym_landau_1991}.  They are ubiquitous to many physical systems~\cite{bardeen_effective_1967,fabrocini_3_1998,franchini_polarons_2021,tajima_intersections_2024}, and also serve as testbeds for studying more complex many-body scenarios~\cite{mahan_many-particle_2000}. Recently, the study of polarons has been revitalized due to their experimental realization with ultracold atomic mixtures~\cite{massignan_polarons_2014,baroni_quantum_2024,grusdt_impurities_2025,massignan_polarons_2025}. They offer a highly controllable setting~\cite{chin_feshbach_2010}, allowing the observation of many polaron phenomena, such as the polaron-to-molecule transition in fermionic mediums~\cite{schirotzek_observation_2009}. Furthermore, impurities serve as a good platform for probing many-body systems, including their temperatures~\cite{hohmann_single-atom_2016,bouton_single-atom_2020}, densities~\cite{adam_coherent_2022}, and even mediated interactions~\cite{edri_observation_2020,paredes_interactions_2024} and phase transitions~\cite{yan_bose_2020,alhyder_mobile_2022,comaron_quantum_2024}.

A rich platform for studying ultracold atomic impurities is optical lattices~\cite{bloch_ultracold_2005,lewenstein_ultracold_2012,gross_quantum_2017}. 
A decade ago, spin impurities interacting with a bosonic bath in a lattice chain were realized experimentally~\cite{fukuhara_quantum_2013}, motivating many theoretical studies of Bose lattice polarons in recent years~\cite{massel_dynamics_2013,dutta_variational_2013,keiler_state_2018,pasek_induced_2019,sarkar_interspecies_2020,keiler_doping_2020,caleffi_impurity_2021,yordanov_mobile_2023,dominguez-castro_bose_2023,zeng_emergent_2023,ding_polarons_2023,colussi_lattice_2023,santiago-garcia_collective_2023,isaule_bound_2024,gomez-lozada_bosefermi_2025,dominguez-castro_polarons_2024,santiago-garcia_lattice_2024,alhyder_lattice_2024,christ_operator_2024}. These include studies of polarons across the superfluid (SF) to Mott-insulator (MI) transition~\cite{fisher_boson_1989,freericks_phase_1994,greiner_quantum_2002}, where quantum critical effects become important. In this direction, it has been shown that the polaron properties are very sensitive to such transition~\cite{colussi_lattice_2023,alhyder_lattice_2024}. 

Particularly relevant is the study of polarons confined in harmonic traps~\cite{dehkharghani_quantum_2015,nakano_bose-einstein-condensate_2017,takahashi_bose_2019,mistakidis_quench_2019,mistakidis_dissipative_2019,mistakidis_effective_2019,mistakidis_radiofrequency_2021,rojo-francas_few_2024,pascual_temperature-induced_2024}, as they better simulate experimental conditions~\cite{bloch_ultracold_2005}. In the case of bosons trapped in optical lattices, harmonic confinement produces neighboring SF and MI domains~\cite{jaksch_cold_1998,batrouni_mott_2002,batrouni_canonical_2008,rigol_state_2009}, which adds a new layer for examining unexplored polaron physics. Impurities could also probe these domains and pave the way to study more complicated atomic mixtures in harmonically confined optical lattices.

\begin{figure}[b!]
	\includegraphics[width=\columnwidth]{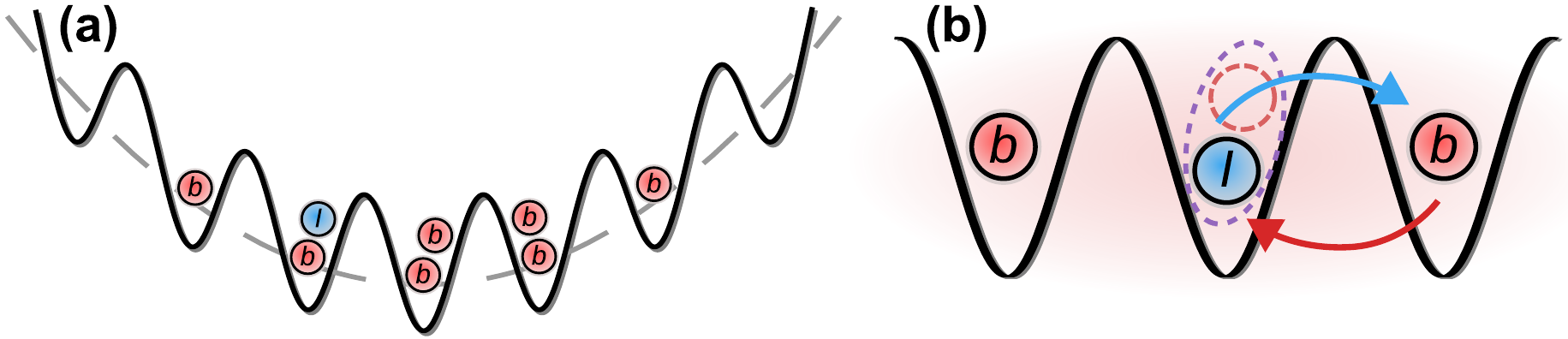}
	\caption{Illustration of the system (a) and of the counterflow state (b) with bath's bosons $b$ and one impurity $I$.}
    \label{sec:intro;eq:illustration}
\end{figure}

In this letter, we consider a mobile impurity interacting repulsively with a bath of bosons, all trapped in a one-dimensional optical lattice. As depicted in Fig.~\ref{sec:intro;eq:illustration}(a), we add an external harmonic potential to study the effect of such a confinement. We examine the system across baths with SF and MI domains and reveal the appearance of a counterflow phase with anti-pair order. Supercounterflows have been recently realized experimentally in binary MIs~\cite{zheng_counterflow_2025} after being predicted for many years~\cite{kuklov_counterflow_2003,altman_phase_2003,kuklov_superfluid-superfluid_2004,hubener_magnetic_2009,hu_counterflow_2009,menotti_detection_2010,nakano_finite-temperature_2012,lingua_demixing_2015,de_forges_de_parny_magnetic_2021}. Thus, this work shows that counterflows can form for high atomic imbalances. Importantly, the impurity and bath become correlated over large distances, showing that it is not a localized effect.
We characterize the properties of the counterflow state and present an analytical model that captures the main physics solely based on the formation of impurity-hole pairs [see  Fig.~\ref{sec:intro;eq:illustration}(b)]. With this model, and by noting that the impurity behaves as a free particle in an infinite square well, we can explain the long-distance behavior of the anti-pair correlator.

\begin{figure*}[t!]
	\includegraphics[width=\textwidth]{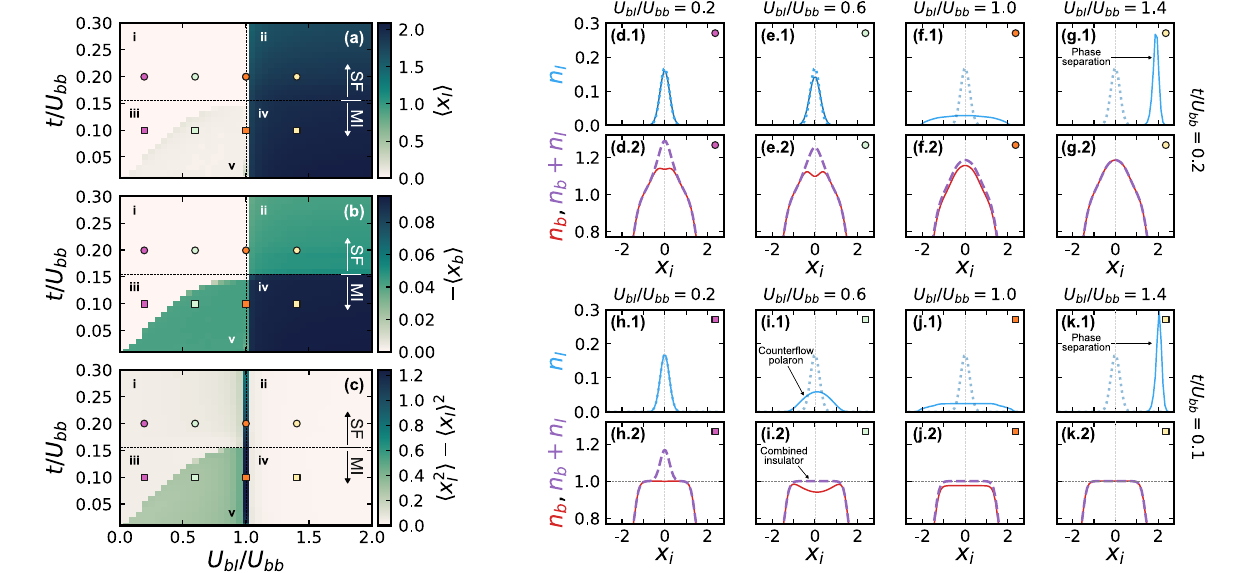}
	\caption{(a--c): Average position of the impurity (a) and of the bath (b), and size of the impurity's cloud (c) as a function of $t/\Ubb$ and $\UbI/\Ubb$. (d--k):  Impurity's [solid blue lines in (d--k.1)], bath's [solid red lines in (d--k.2)], and combined bath and impurity [dashed purple lines in (d--k.2)] profiles as a function of $x_i$ for the interactions indicated on top and right of the panels. The dotted blue lines in (d--k.1) correspond to the profile of a free impurity ($\UbI=0$).}
    \label{sec:phasediagram;eq:xprofiles}
\end{figure*}

\textbf{\textit{Model.}}-- 
We consider a binary Bose-Hubbard model for $N_b$ bath bosons ($\sigma=b$) and one impurity ($\sigma=I$)
\begin{align}
    \Hop =& - t\sum_{i,\sigma}\left(\aopd_{i,\sigma}\aop_{i+1,\sigma}+\textrm{h.c.}\right)+\Vho\sum_{i,\sigma}i^2\nop_{i,\sigma}\nonumber\\
    &+\frac{\Ubb}{2}\sum_{i} \nop_{i,b}\left(\nop_{i,b}-1\right)+\UbI\sum_{i}\nop_{i,b}\nop_{i,I},
    \label{sec:model;eq:H}
\end{align}
where $\aop_{i,\sigma}$ ($\aopd_{i,\sigma}$) annihilates (creates) a boson of species $\sigma=b,I$ at site $i$ and $\nop_{i,\sigma}=\aopd_{i,\sigma}\aop_{i,\sigma}$ is the number operator. The trap's center lies at $i=0$, and thus the sites run between $\pm(M-1)/2$, with $M$ the number of sites. The first term describes the tunneling of atoms, with equal tunneling coupling $t>0$ for both species. The second term describes the harmonic trap, where $\Vho$ dictates its strength.
The third and fourth terms describe the on-site boson-boson and bath-impurity interactions, respectively. Both interactions are repulsive ($\Ubb,\UbI>0$). 

The properties of Bose lattice polarons are dictated by $t/\Ubb$ and $\UbI/\Ubb$~\cite{colussi_lattice_2023,yordanov_mobile_2023}. However, the harmonic trap adds an additional scale, $\xi=d\sqrt{t/\Vho}$, where $d$ is the lattice spacing, which defines the bath's characteristic density $\tilde{\rho}_b=d N_b/\xi$. The properties of the harmonically confined Bose-Hubbard model become invariant for constant $\tilde{\rho}_b$~\cite{batrouni_canonical_2008}. Thus, this work examines the system for a varying range of $t/\Ubb$ and $\UbI/\Ubb$ but for a fixed density of $\tilde{\rho}_b=3.58$. This choice results in a bath that is completely SF for $t/\Ubb \gtrsim 0.155$ and that has a central MI domain of unity filling for \emph{all} $t/\Ubb \lesssim 0.155$~\cite{batrouni_canonical_2008}. We perform numerical simulations using the density matrix renormalization group (DMRG) method~\cite{white_density_1992,schollwock_density-matrix_2005,schollwock_density-matrix_2011,orus_practical_2014} for $N_b=40$ and $M=60$. Nevertheless, our results are valid for any large $N_b$ due to the system's invariance. See the supplemental material for additional details.

\textbf{\textit{Phase diagram and density profiles.}}-- To examine the behavior of the system under different interaction strengths, we compute the density profiles $n_{\sigma}(i)=\langle \nop_{i,\sigma} \rangle$. We study the profiles as a function of the rescaled length $x_i=i\,d/\xi$~\cite{batrouni_canonical_2008}. To characterize the phase diagram, we calculate the average position $\langle x_\sigma\rangle$ of each species and the size of the impurity's cloud $\langle x^2_I \rangle$-$\langle x_I\rangle^2$, where
\begin{equation}
    \langle x_\sigma \rangle = \frac{1}{N_\sigma}\sum_{i}x_i\, n_\sigma(i),\qquad \langle x^2_I \rangle =\sum_{i} x_i^2\, n_I(i)\,.
\end{equation}
We show average positions and clouds' sizes in Fig.~\ref{sec:phasediagram;eq:xprofiles}(a--c), which show well-defined phases. Profiles for a representative set of interactions are shown in panels (d--k).

We first examine completely SF baths [$t/\Ubb\gtrsim 0.155$]. When $\UbI<\Ubb$ [region i in panels (a--c)], we observe that $\langle x_\sigma\rangle=0$ ($\sigma=b,I$), and thus both the bath and impurity are centered in the trap. This is illustrated by the profiles (d) and (e), where the bath develops a small decrease in its density around $x_i=0$ due to the impurity's repulsion. This decrease is only possible due to the compressibility of the SF bath. The extension of the impurity also shows a very small increase with $\UbI$, albeit difficult to observe in panel (c). 
Then, for $\UbI=\Ubb$, the system behaves as an SF gas with $N_b+1$ bosons. This is illustrated by panels (f), where the impurity develops an enlarged profile, as captured by the large values in panel (c) for $\UbI=\Ubb$.  Finally, when $\UbI>\Ubb$ [region ii], the impurity and bath develop finite average positions with opposite signs and a very large $\langle x_I\rangle$, signaling a \emph{phase separation} between the species. This immiscibility is illustrated by panels (g), where the impurity is repelled to the right border of the trap~\footnote{Note that the impurity localizes at only one border of the trap instead of developing a superposition on both sides. This is a feature of the DMRG calculations. In contrast, exact diagonalization calculations~\cite{raventos_cold_2017} do recover a symmetric solution with peaks at both borders.}. The bath also moves slightly to the left, as illustrated by panel (b) where $\langle x_b\rangle\approx -0.05$. Note that in all cases the combined profiles $n_b+n_I$ [dashed purple lines in panels (d--g).2] show a Gaussian-like profile. Overall, the system's behavior is similar to what is observed in non-lattice harmonically confined Bose polarons~\cite{mistakidis_quench_2019,mistakidis_radiofrequency_2021}.

The situation changes when considering baths with an MI domain [$t/\Ubb\lesssim 0.155$]. For small $\UbI$ [region iii], once again both the bath and impurity are centered in the trap ($\langle x_\sigma\rangle=0$). However, the profiles remain unchanged with $\UbI$. Indeed, in panel (h.1), the non-interacting profile (dotted curve) overlaps with the interacting one (solid curve). In turn, the bath [panel (h.2)] shows its MI domain at the center, remaining undisturbed by the impurity. This behavior is due to the incompressibility of the MI domain.  
Turning to the case where $\UbI=\Ubb$ [panels (j)], the system again behaves as a one-component one with $N_b+1$ bosons, and so the impurity's profile is that of one particle in such a system. Then, for large $\UbI>\Ubb$ [region iv], the system undergoes a phase separation (large $\langle x_I\rangle$), as illustrated by panels (k). Due to its incompressibility, the MI bath centers at $\langle x_b\rangle\approx -0.09$, thus moving further away than an SF bath. Finally, an additional phase with counterflow order appears for intermediate interactions (region v), which we examine next. 

\begin{figure}[t]
\centering
\includegraphics[width=\columnwidth]{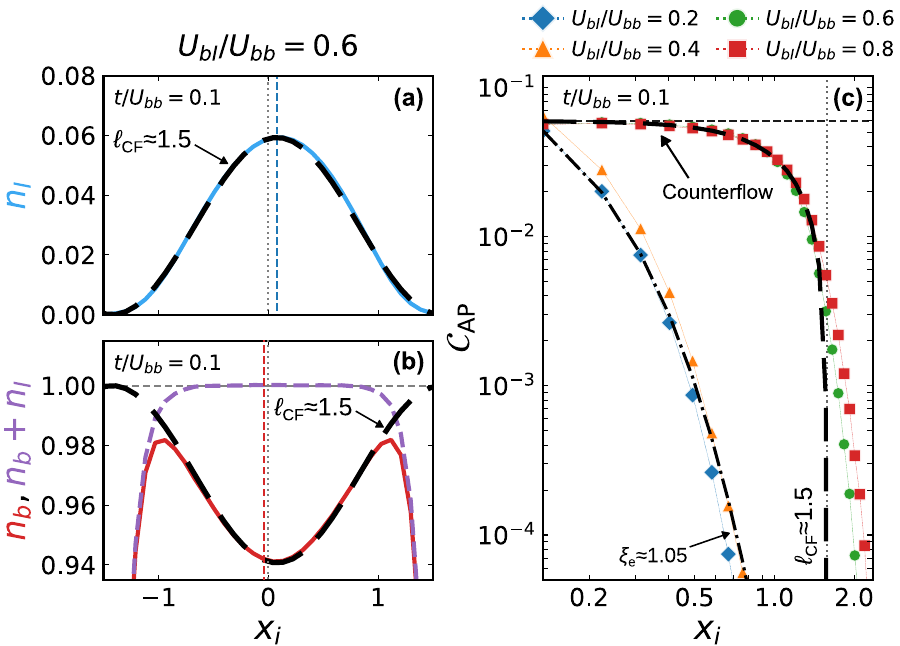}
\caption{Impurity's (a), bath's [solid red line in (b)] and combined [dashed purple line in (b)] profiles, and anti-pair correlator (c) as a function of $x_i$ for $t/\Ubb=0.1$. Panels (a) and (b) consider $\UbI/\Ubb=0.6$, while panel (c) considers the values indicated in the legend. The long-dashed black lines in (a--c) are functions (\ref{sec:CF;eq:nICF}) and (\ref{sec:CF;eq:CSCFansatz}), and the dash-dotted line in (c) is an exponential fit for $\UbI/\Ubb=0.2$.}
\label{sec:CF;fig:CF}
\end{figure}

\textbf{\textit{Counterflow phase.}}-- Region v of Fig.~\ref{sec:phasediagram;eq:xprofiles}(a--c) develops notorious changes in the profiles, which are illustrated by Fig.~\ref{sec:phasediagram;eq:xprofiles}(i). In panel (i.1), the interacting impurity's profile (solid line) deviates strongly from the non-interacting one (dashed line), developing an enlarged profile. The latter is captured by the larger cloud's size shown in Fig.~\ref{sec:phasediagram;eq:xprofiles}(c). 
Both species also develop a finite but small $\langle x_\sigma\rangle$ [panels (a) and (b)], difficult to observe in the profiles. Importantly, the bath decreases its density around the trap's center instead of showing a domain of constant unity filling. Therefore, the stronger bath-impurity repulsion destroys the MI domain, making the bath compressible. For clarity, the profiles can be appreciated in greater detail in Fig.~\ref{sec:CF;fig:CF}(a,b).

A striking feature of this phase is that the sum of the profiles [dashed purple lines in Figs.~\ref{sec:phasediagram;eq:xprofiles}(i.2) and~\ref{sec:CF;fig:CF}(b)] form a domain of unity filling, signaling an insulating state. Thus, the bath and impurity form a combined insulator phase, even though independently, each species does not show features of an MI domain. Indeed, the independent profiles do not show any plateau of constant density, only the sum of the two does. This behavior can be explained by the appearance of a counterflow state.

Long-range counterflow order is characterized by a slow algebraic decay of the anti-pair correlator~\cite{hu_counterflow_2009}
\begin{equation}
    \CAP(i)=\langle \aop_{0,b}\aopd_{0,I}\aopd_{i,b}\aop_{i,I}\rangle.
\end{equation}
In contrast, a rapid exponential decay indicates the absence of counterflow. As a function of $x_i$, the latter can be written as $\propto \exp(-|x_i|/{\xi_{\text{e}}})$. We show both correlators for a representative set of interactions in Fig.~\ref{sec:CF;fig:CF}(c).

$\CAP$ decays exponentially in all regions i to iv of the phase diagram. In Fig.~\ref{sec:CF;fig:CF}(c), this rapid decay is illustrated by the results for $\UbI/\Ubb=0.2$ and $\UbI/\Ubb=0.4$ [region iii]. In contrast, for $\UbI/\Ubb=0.6$ and $\UbI/\Ubb=0.8$ [region v], $\CAP$ shows a noticeable slower  decay. Moreover, the correlator is approximately constant for $x_i\lesssim 0.5$. We have found that this slow decay appears suddenly in region v, and thus, we can conclude that such a region corresponds to a counterflow phase. These results confirm that counterflow order can appear in highly imbalanced configurations, thus not being limited to balanced or near-balanced mixtures.

In a counterflow state, each species tunnels in opposite directions [see Fig.~\ref{sec:intro;eq:illustration}(b)]. Importantly, counterflows are insulators, even though the particles can still move~\cite{zheng_counterflow_2025}. This correlated insulator state explains the appearance of the combined profile of unity filling $n_b+n_I=1$. However, while balanced counterflows show individual profiles with plateaus of constant density~\cite{hu_counterflow_2009}, in our system, the bath and impurity do not show such plateaus, as discussed. 
This is unexpected, as one would naively predict plateaus with filling of $N_b/(N_b+1)$ and $1/(N_b+1)$. 
These unusual profiles make the picture of an insulator particularly counterintuitive, as the impurity prefers to flow at the center of the lattice while the bath shows density peaks at $|x_i|\approx 1$. Nevertheless, the behavior of $\CAP$ confirms the appearance of a counterflow.

The slow decay of $\CAP$ and the large extension of $n_I$ also demonstrate that the counterflow state forms over large distances. Therefore, this single impurity suddenly forms particle-hole pairs with the whole MI domain, indicating that it is not a localized effect.

The shape of the impurity's profile is particularly interesting. In the non-counterflow phases, $n_I$ are Gaussian-like functions, where their widths are dictated by the harmonic trap.  In contrast, we have found that the counterflow impurity's profile takes the form
\begin{equation}
    \nICF(x_i)=n_{I}^{(0)}\cos^2\left(\pi (x_i-\langle x_I\rangle)/\lCF\right),
    \label{sec:CF;eq:nICF}
\end{equation}
where $n_{I}^{(0)}$ is the maximum value of the profile and $\lCF=2/(\xi n_{I}^{(0)})$ so that $\sum_i \nICF=1$. Function~(\ref{sec:CF;eq:nICF}) is shown in Fig.~\ref{sec:CF;fig:CF}(a) by the long-dashed line. In turn, $1-\nICF$ is shown in Fig.~\ref{sec:CF;fig:CF}(b), which agrees with $n_b$ almost perfectly for $|x_i| \lesssim 1$. We have found a value of $\lCF\approx 3$ for $t/\Ubb=0.1$ and $\UbI/\Ubb=0.6$, even though $\lCF$ is roughly constant in the whole counterflow phase. 

Eq.~(\ref{sec:CF;eq:nICF}) describes the ground-state density probability of a free particle in an infinite square well of width $\lCF$. This can be understood as an effective behavior. The appearance of this length scale $\lCF$ stems from the strong correlations between the impurity and the bath, which dominate over the effect of the harmonic trap. 

To further understand the counterflow state, we propose a simple model based on an impurity-hole pair in a chain with $\MCF$ sites. We consider
\begin{equation}
    \MI=\prod_i \aopd_{i,b}|\emptyset \rangle,\qquad \iIH = \aopd_{i,I}\aop_{i,b}\MI,
\end{equation}
where $\MI$ and  $\iIH$ are the MI and impurity-hole states, respectively. The counterflow wavefunction is then
\begin{equation}
    \PsiCF=\sum_i \alpha_i \iIH,
\end{equation}
where $\sum_i |\alpha_i|^2=1$. With this wavefunction, the impurity's profile reads $n_I(i)=|\alpha_i|^2$. Importantly, after some algebra, one obtains that the bath behaves as $n_b(i)=1-|\alpha_i|^2$, recovering the result $n_b+n_I=1$. In addition, the anti-pair correlator takes the form $\CAP(i)=\alpha_0^*\alpha_i$. By using Eq.~(\ref{sec:CF;eq:nICF}), one obtains a trigonometric dependence
\begin{equation}
    \CAP(i)=n_I^{(0)}\cos(i\pi/\MCF),
    \label{sec:CF;eq:CSCFansatz}
\end{equation}
which predicts a slow decay of $\CAP$. The long-dashed line in Fig.~\ref{sec:CF;fig:CF}(c) shows function~(\ref{sec:CF;eq:CSCFansatz}) with the constants taken from Eq.~(\ref{sec:CF;eq:nICF}). Note that we have used $\lCF=d \MCF/\xi$ and $x_i\to(x_i-\langle x_I\rangle)$. The cosine gives a near-perfect agreement, and thus we can conclude that this simple impurity-hole model nicely explains the behavior of $\CAP$. Here we mention that while a Gaussian fit can qualitatively describe the profile of the impurity, its exponential nature cannot describe the slow decay of $\CAP$.

Finally, the MI-counterflow transition studied here does not appear in lattices with no harmonic trap. In such lattices, an MI phase is only possible for integer filling $N_b=\nu M$, with $\nu$ an integer. However, a counterflow state requires that $N_b+N_I=\nu$. Thus, the harmonic trap is required to find the rich physics presented here.

\begin{figure}[t]
\centering
\includegraphics[width=\columnwidth]{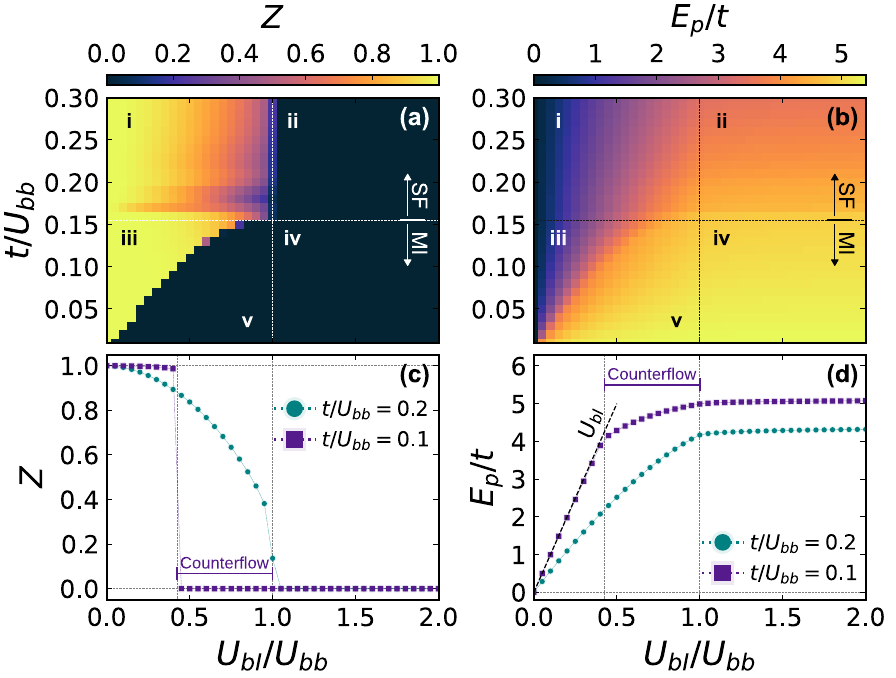}
\caption{Polaron residue (a,c) and energy (b,d) as a function of $\UbI/\Ubb$ (all panels) and $t/\Ubb$ (top panels). The vertical dashed lines at $\UbI/\Ubb\approx 0.4$ in the bottom panels indicate the transition to the counterflow phase for $t/\Ubb=0.1$. }
\label{sec:polaron;fig:polaron}
\end{figure}

\textbf{\textit{Polaron properties.}}-- We now examine experimentally measurable polaron properties across the phase diagram. We calculate the polaron residue~\cite{guenther_mobile_2021}
\begin{equation}
    Z(\UbI)=|\langle \Psi(\UbI=0)|\Psi(\UbI)\rangle|^2\,,
\end{equation}
where $|\Psi\rangle$ is the ground-state wavefunction. The residue measures the prevalence of the free impurity state in the interacting one. We also calculate the polaron energy
\begin{equation}
    E_p(\UbI)=E(\UbI)-E(\UbI=0),
\end{equation}
where $E$ are ground-state energies, and thus $E_p$ measures the energy required to add the impurity to the bath. We show both quantities in Fig.~\ref{sec:polaron;fig:polaron}.

Firstly, the residue in panel (a) strongly marks the transition between the miscible regions (i and iii) and the other phases (ii, iv, and v). A vanishing residue for $\UbI>\Ubb$ is expected in immiscible configurations~\cite{mistakidis_quench_2019}. However, the residue also vanishes in the counterflow phase. This orthogonality suggests an \emph{orthogonality catastrophe} in the thermodynamic limit when counterflow order appears, indicating the breaking of the traditional polaron picture. We can understand this orthogonality by realizing that, due to the appearance of sites with holes, the Fock states that describe a counterflow configuration become orthogonal to those of an unperturbed MI bath.

Interestingly, in SF baths [region i] the residue decreases smoothly with increasing $\UbI$ [see circle markers in (c)], similar to other scenarios where polarons show orthogonality. In contrast, in baths with MI domains, the residue remains constant and equal to one for small $\UbI$ [region ii], which we attribute to the incompressibility of the MI bath. Then, the residue vanishes rapidly at the transition to the counterflow phase [region v]. This behavior is well-illustrated by the square markers in panel (c). Here we stress that while the numerical calculations do not show a fully discontinuous residue, the discontinuity becomes more abrupt with a larger number of particles.
This orthogonality is another feature of the counterflow phase, which could be probed experimentally via radio-frequency spectroscopy~\cite{shashi_radio-frequency_2014}. 

Finally, the polaron energy [panel (b)] also shows distinct behaviors at the different phases, albeit with smooth transitions and no discontinuities. In all cases, $E_p$ increases with $\UbI$, to then saturate for large $\UbI$ [see panel (d)], agreeing qualitatively with related studies~\cite{pasek_induced_2019,yordanov_mobile_2023}.  Importantly, in baths with MI domains [see square markers in (d)], the polaron energy initially increases as $E_p=\UbI$ because the impurity only interacts with one boson per site. Then, in the counterflow phase, the polaron energy deviates from such dependency, showing a slower increase with $\UbI$ due to the change in the kinetic and interacting energies. These dependencies could be measured experimentally~\cite{jorgensen_observation_2016}, providing an additional option to probe the counterflow phase.

\textbf{\textit{Conclusions.}}-- We have shown that a mobile impurity interacting repulsively with a bosonic bath in a harmonically confined optical lattice shows a very rich behavior, including the appearance of an unconventional counterflow phase. This counterflow is characterized by the appearance of a combined domain of unity filling with long-range anti-pair correlations. Contrary to intuition, we found that in this phase the impurity behaves as a free particle moving in an effective infinite square well. The counterflow phase transition also features a sudden orthogonality between the interacting and non-interacting ground states. These features could be measured in current experiments with highly-imbalanced ultracold atomic mixtures, and also be used to probe the SF-MI transition in harmonically confined optical lattices. The findings on counterflows could be further explored theoretically with imbalanced mixtures~\cite{valles-muns_quantum_2024} and bipolarons~\cite{ding_polarons_2023,isaule_bound_2024}, and also experimentally following the recent realization of supercounterflows~\cite{zheng_counterflow_2025}. Additionally, the revealed counterflow state of impurities could be used to probe and simulate broader anti-pair behaviors of polarons, such as in analog models~\cite{mardonov_gravitating_2022}.\\

\textit{Acknowledgments.}-- 
We thank J. Martorell for the useful discussions.
F.I. acknowledges funding from ANID through FONDECYT Postdoctorado No. 3230023. This work has been funded by Grant PID2023-147475NB-I00 funded by MICIU/AEI/10.13039/501100011033 and FEDER, UE, by grants 
2021SGR01095 from Generalitat de Catalunya, and by 
Project CEX2019-000918-M of ICCUB (Unidad de Excelencia María 
de Maeztu). A.R.-F. acknowledges funding from MICIU through grant FPU20/06174. F.I. and L.M.-M. acknowledge additional funding from Instituto de Física UC. 
DMRG calculations were performed using the \texttt{ITensor} library~\cite{fishman_itensor_2022}.


\bibliography{biblio}

%

\appendix
\clearpage
\widetext{

\section{Supplemental Material for:\texorpdfstring{\\}{ } Counterflow of lattice polarons in harmonically confined optical lattices}

\section{Numerical details}
\label{sec:calculations}

To study our system we employ the DMRG method~\cite{schollwock_density-matrix_2005,schollwock_density-matrix_2011}, enabling us to perform highly accurate calculations for systems with a large number of sites and particles. We perform the DMRG calculations using the publicly available \texttt{ITensor} library~\cite{fishman_itensor_2022}. We work with a maximum number of $N_\text{max}=5$ bosons per site, which we have checked is large enough to achieve convergence for the parameters of $t/\Ubb$ and $\Vho/t$ used in this work. Our calculations consider a maximum bond dimension of $\chi=1000$~\cite{orus_practical_2014} and a decreasing noise in the first ten DMRG sweeps to avoid getting stuck at local minima~\cite{white_density_1992}. The solution is extracted after two subsequent sweeps have a percentual energy difference smaller than 10$^{-10}$. If such convergence is not achieved after 20 sweeps, the DMRG process is started again with a larger Krylov space. We refer to the official \texttt{ITensor} documentation for a detailed discussion on the convergence of the DMRG method [\href{https://itensor.github.io/ITensors.jl/dev/faq/DMRG.html}{https://itensor.github.io/ITensors.jl/dev/faq/DMRG.html}]. We also refer to Ref.~\cite{valles-muns_quantum_2024} for a detailed discussion on the convergence for a related two-component Bose-Hubbard model. 

Following Ref.~\cite{batrouni_mott_2002}, the calculations shown in the main text employ $\Vho=0.008 t$ and $N_b=40$, resulting in the chosen characteristic density of approximately $\tilde{\rho}_b= 3.58$. 
For benchmarking, we also show in this supplemental material results from DMRG calculations for the same density $\tilde{\rho}_b$ but with $\Vho=0.0205 t$ and $N_b=25$. For both parameter choices, including the main text, we consider lattices with $M=N_b+20$ sites. These large lattice sizes enable us to avoid any artificial border effect.  To aid the convergence, the DMRG results shown in this work also added a very small tilt to the lattice so that, if the profiles develop a finite average value, the impurity prefers to localize at the right, while the bath prefers to localize at the left. However, we have also performed calculations without such a tilt, resulting in identical results, but with the impurity (bath) localizing randomly at either the left (right) side or vice versa.

In this supplemental material, and for additional benchmarking, we also show some exact diagonalization (ED)~\cite{raventos_cold_2017} calculations on small lattices with $\Vho=0.2 t$, $N_b=8$, and $M=14$. While such small lattices show smooth crossovers instead of well-defined transitions, they provide a good cross-check for our DMRG results, enabling us to rule out any convergence issue in our simulations. 

\section{Mott insulator--superfluid transition and bath parameters}
\label{sec:bath}
Firstly, we examine the behavior of the bath in the absence of the impurity to illustrate the onset of the superfluid (SF) and Mott insulator (MI) domains and find the phase transition point (for a detailed examination see Refs.~\cite{batrouni_mott_2002,batrouni_canonical_2008}). This is modeled by the one-component Bose-Hubbard Hamiltonian with a harmonic confinement
\begin{equation}
    \Hop = - t\sum_{i}\left(\aopd_{i}\aop_{i+1}+\textrm{h.c.}\right)+\Vho\sum_{i} i ^2\nop_{i}+\frac{U}{2}\sum_{i} \nop_{i}\left(\nop_{i}-1\right),
\label{sec:bath;eq:BH}
\end{equation}
where $\aop_{i}$ ($\aopd_{i}$) annihilates (creates) a boson at site $i=-(M-1)/2,\dots,(M-1)/2$ and $\nop_{i}=\aopd_{i}\aop_{i}$ is the number operator. Note that this section drops the $b$ subscript, as we only consider a one-component system.
We consider systems with characteristic densities of approximately $\tilde{\rho}=d N/\xi= 3.58$. As previously mentioned, we perform DMRG calculations for $N=40$ and $N=25$ bosons, and ED calculations for $N=8$ bosons. Fig.~\ref{sec:bath;fig:BH}(a--c) shows the density profile $  n(i)=\langle \nop_i\rangle$ for selected parameters. The results are shown as a function of the centered rescaled length $x_i=i\,d/\xi$.

Panel (a) illustrates the behavior of the bosons deep in the SF regime, with a near-Gaussian density profile. An SF system near the transition point is shown in panel (b), where the profile shows a deformed Gaussian-like form. Finally, the appearance of the MI domain is illustrated by panel (c), where the density profile shows a large domain of constant density $n(i)=1$. In all cases, the DMRG curves for $N=25$ are almost completely underneath those for $N=40$, thus providing nearly identical results. This exemplifies the near-universality of the model for constant $\tilde{\rho}$. The ED calculations also show similar profiles, only showing minor differences with the DMRG curves. Nevertheless, it is striking that ED can capture the overall shape of the profiles with such a small number of particles and sites.

\begin{figure}[t]
\centering
\includegraphics[width=\textwidth]{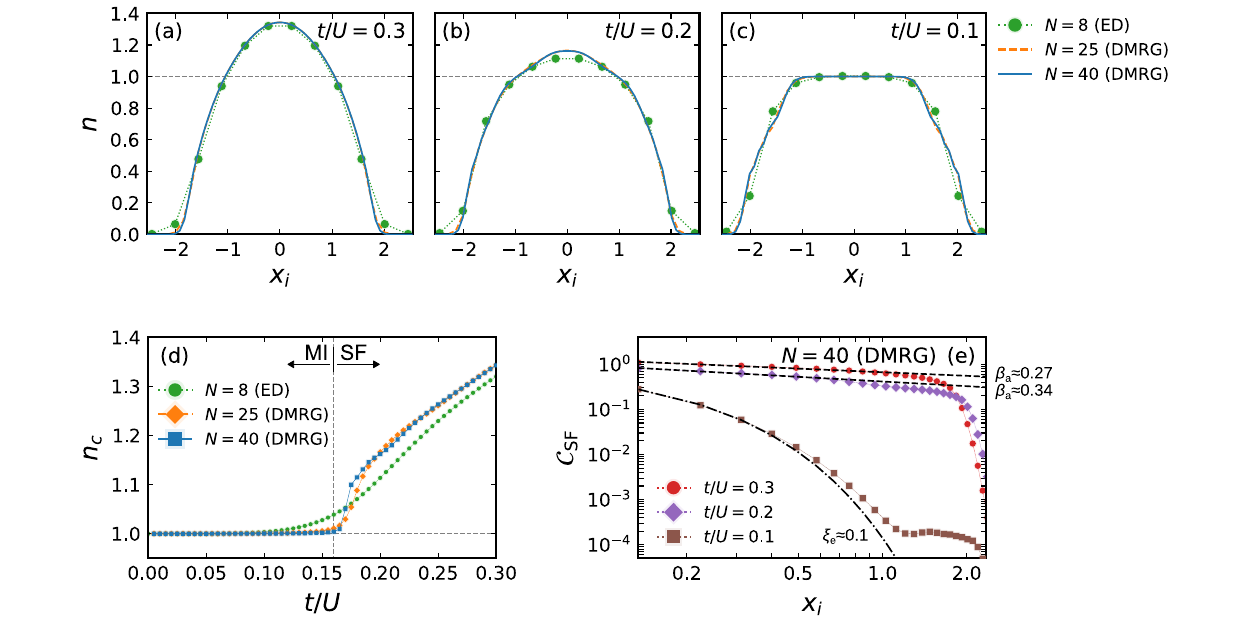}
\caption{Results for the one-component Bose-Hubbard model~(\ref{sec:bath;eq:BH}) with $\tilde{\rho}= 3.58$. (a--c): Density profiles $n$ as a function of  $x_i$ for the values of $t/U$ indicated in each panel. (d): Central density $n_c=n(i=0)$  as a function of $t/U$. Panels (a--d) include ED calculations with $N=8$ and $\Vho=0.2t$, DMRG calculations with $N=25$ and $\Vho=0.0205t$, and DRMG calculations with $N=40$ and $\Vho=0.008t$, as indicated in the legends. (e): One-body correlator as a function of $x_i$ for the values of $t/U$ indicated in the legend. Panel (e) only shows DMRG calculations with $N=40$. The dashed black lines are algebraic fits for $t/U=0.3$ and $t/U=0.2$, and the dash-dotted black line is an exponential fit for $t/U=0.3$.}
\label{sec:bath;fig:BH}
\end{figure}

To pinpoint the phase transition point, Fig.~\ref{sec:bath;fig:BH}(d) shows the values of the densities at the center of the lattice as a function of $t/U$. The central density $n_c$ decreases with decreasing tunneling $t/U$. For large $t/U$ the central density has a value $1<n_c<2$, indicating that all the bosons are in an SF state. However, for small $t/U$ the central density $n_c$ becomes constant and equal to one, showing the onset of the Mott domain for strong interactions. The DMRG calculations show a fairly well-defined phase transition at $t/U\approx 0.155$ (vertical dashed gray lines) with the saturation of $n_c$ to unity. In contrast, the ED results show a continuous crossover, as expected from such few-body calculations. Nevertheless, we can conclude that $t/U\approx 0.155$ corresponds to the SF-to-MI transition point for $\tilde{\rho}= 3.58$, agreeing with that reported in Ref.~\cite{batrouni_canonical_2008}.

Finally, to further characterize the phases, in Fig.~\ref{sec:bath;fig:BH}(e) we show the long-distance behavior of the one-body correlator
\begin{equation}
    \CSF(i)=\langle \aop_{0}\aopd_{i}\rangle.
\end{equation}
In the SF phase, this correlator decays algebraically, that is as $\mathcal{C}_{\text{SF}}(x_i)\propto 1/|x_i|^{\beta_{\text{a}}}$ when written as a function of $x_i$. Such a slow decay indicates long-range correlations. In contrast, it decays exponentially as $\mathcal{C}_{\text{SF}}(x_i)\propto \exp(-|x_i|/{\xi_{\text{e}}})$ in the MI phase. It is straightforward to see that for $t/U=0.3$ and $t/U=0.2$ the correlator decays algebraically for $x_i\lesssim 2$, as illustrated by the dashed fits. In turn, for $t/U=0.1$ the correlator decays exponentially within the MI domain ($x_i\lesssim 1.0$) as illustrated by the dash-dotted fit.

\section{Comparison between DRMG and ED calculations}

We have performed several comparisons between the different DMRG simulations and found near-identical results. The ED calculations also show qualitatively similar results. To briefly illustrate this, in Fig.~\ref{sec:benchmark;fig:benchmark} we show the polaron residue and energy, as defined in the main text. In addition, we show the average distance $\langle \rbI \rangle$ between the impurity and the bosons in the bath to easily compare the distribution of the particles obtained from the DMRG and ED calculations. Within DRMG, these are calculated from
\begin{equation}
    \langle \rbI \rangle = \frac{1}{N_b}\sum_{i,j}|x_i-x_j|n_I(i)n_b(j),
\end{equation}
while from ED they are calculated from each Fock state (see Ref.~\cite{isaule_bound_2024} for details). Note that we normalize $\rbI$ by $\xi$ to compare the distances correctly.

\begin{figure}[t]
\centering
\includegraphics[width=\textwidth]{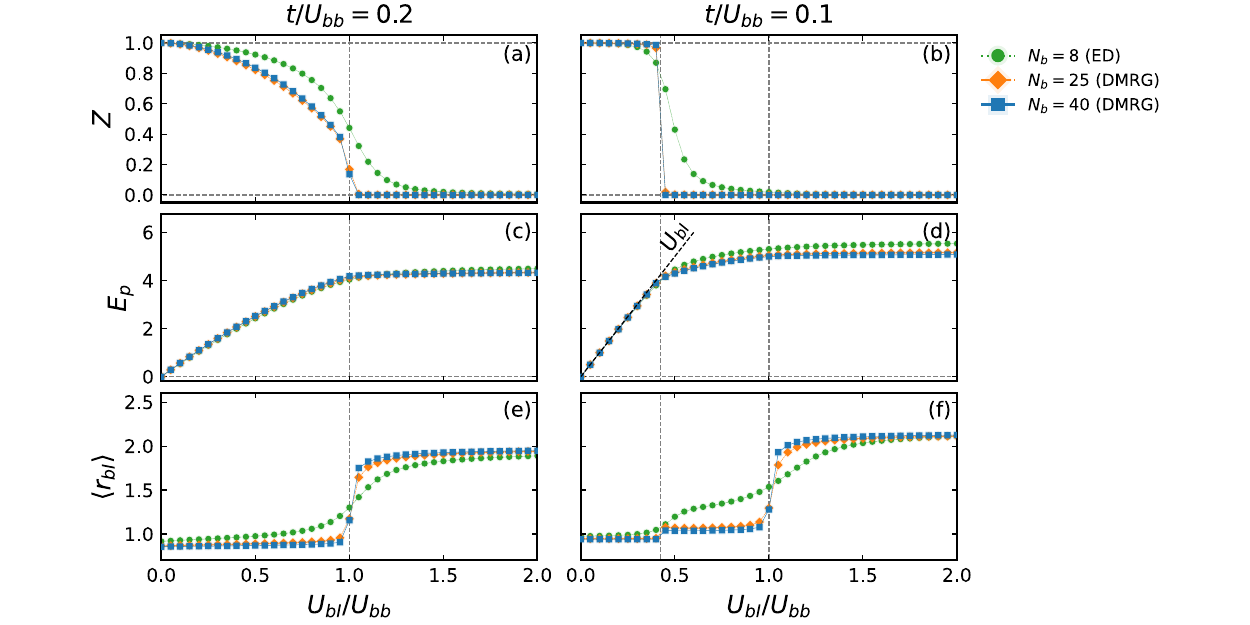}
\caption{Polaron residue  (a,b), polaron energy  (c,d), and the average distance between the bosons and the impurity  (e,f) as a function of $\UbI/\Ubb$ for $t/\Ubb=0.2$ (left panels) and $t/\Ubb=0.1$ (right panels). The panels include ED calculations with $N=8$ and $\Vho=0.2t$, DMRG calculations with $N=25$ and $\Vho=0.0205t$, and DRMG calculations with $N=40$ and $\Vho=0.008t$, as indicated in the legends. In the right panels, the region $0.425\lesssim \UbI/\Ubb\leq1$ corresponds to the counterflow phase.}
\label{sec:benchmark;fig:benchmark}
\end{figure}

As mentioned, both DMRG calculations provide almost identical results. In turn, the ED results agree qualitatively, but show smooth crossovers, as expected. In particular, the average distance shows a large increase around $\UbI=\Ubb$ in all cases, showing that all calculations predict a phase separation at the same point. Similarly, $\rbI$ shows a small but noticeable increase in the counterflow phase. This indicates that the bosons and the impurity do not reside at the same site despite tunneling within the same region, as expected from a counterflow state. 

Other quantities also show similar agreements. Thus, overall, all the calculations agree, confirming the robustness of our results.

\section{Impurity-hole model}

Here we provide a more detailed presentation of the impurity-hole model. As stated in the main text, we start with an MI state for the bath, $\MI=\prod_i \aopd_{i,b}|\emptyset\rangle$, in a chain with $\MCF$ sites. An elementary impurity-hole state is then defined by $\iIH = \aopd_{i,I} \aop_{i,b}|\text{MI}\rangle$, resulting in the wavefunction $\PsiCF=\sum_i \alpha_i \iIH$, where $\sum_i |\alpha_i|^2=1$. Note that this wavefunction does not conserve the number of bosons, and thus is only an approximation for a large number of particles. Naturally, this restriction stems from the consideration of a chain in this simple model. In the full model, the addition of the harmonic trap provides the additional site needed to create the annihilated boson.

To obtain the density of the impurity, we first calculate
\begin{align}
    \nop_{i,I}\PsiCF &=\aopd_{i,I}\aop_{i,I}\sum_j \alpha_j \aopd_{j,I}\aop_{j,b}\MI\nonumber\\
    &=\alpha_i\aopd_{i,I}\aop_{i,b}\MI,
\end{align}
which results in%
\begin{align}
    n_I(i)&=\PsiCFc \nop_{i,I}\PsiCF\nonumber\\
    &=\sum_j \MIc \alpha^*_j \aop_{j,I}\aopd_{j,b}\alpha_i\aopd_{i,I}\aop_{i,b}\MI\nonumber\\
    &=|\alpha_i|^2,
\end{align}
as reported. In turn, for the bath's profile, we have that
\begin{align}
    \nop_{i,b}\PsiCF & = \aopd_{i,b}\aop_{i,b}\sum_j \alpha_j \aopd_{j,I}\aop_{j,b}\MI\nonumber\\
    &=\sum_{j\neq i} \alpha_j \aopd_{j,I}\aop_{j,b}\MI\nonumber\\
    &=\PsiCF-\alpha_i\aopd_{i,I}\aop_{i,b}\MI,
\end{align}
where it was imposed in the second line that $\MI$ supports only one boson per site. To take the bracket of the second term in the third line, we now evaluate
\begin{align}
    \aopd_{i,b}\aop_{i,I}\PsiCF &= \alpha_i\aopd_{i,b}\aop_{i,I}\sum_j \alpha_j \aopd_{j,I}\aop_{j,b}\MI\nonumber\\
    &=\alpha_i \aopd_{i,b}\aop_{i,b}\MI\nonumber\\
    &=\alpha_i\MI.
\end{align}
Thus, one obtains the reported result
\begin{align}
    n_b(i)&=\PsiCFc \nop_{i,b}\PsiCF\nonumber\\
    &=\PsiCFc\left(\PsiCF-\alpha_i\aopd_{i,I}\aop_{i,b}\MI\right)\nonumber\\
    &=1-|\alpha_i|^2.
\end{align}
As stressed in the main text, these results recover the combined filling of $n_b+n_I=1$.

Finally, the anti-pair correlator reads
\begin{align}
    \CAP(i)=&\PsiCFc \aop_{0,b}\aopd_{0,I}\aopd_{i,b}\aop_{i,I}\PsiCF\nonumber\\
    =&\alpha_0^*\alpha_i.
\end{align}
By now imposing that the coefficients $\alpha_i$ are real and positive, from $n_I(i)=|\alpha_i|^2=n_I^{(0)}\cos^2(i\pi/\MCF)$ we get that
\begin{equation}
    \alpha_i=\sqrt{n_I^{(0)}}\cos(i\pi/\MCF).
\end{equation}
With this, the correlator finally takes the form $\CAP(i)=n_I^{(0)}\cos(i\pi/\MCF)$. Here $\MCF=2/n_I^{(0)}$, so that there is only one impurity in the system ($\sum_{i}n_I(i)=1$). Note that this sum of the profile needs to be performed in $i$ instead of the rescaled length $x_i$, explaining the value $\lCF=2/(\xi n_{I}^{(0)})$ reported in the main text.

\end{document}